\documentclass[a4paper,11pt]{article}
\usepackage{pos}

\title{The hadronic decays of charmed mesons at BESIII}
\author*[a]{Hui Li} 
\affiliation[a]{Nankai Univesity,\\
94 Weijin Road, Nankai District, Tianjin, China}

\emailAdd{huili@mail.nankai.edu.cn}

\abstract{
In this presentation, we discuss the recent hadronic decays of charmed mesons observed in the BESIII experiment, specifically focusing on the decay processes $D^+ \to K_S^0 \pi^+ \eta$, $D^0 \to \pi^+ \pi^- \eta$, $D^+ \to \pi^+ \pi^0 \eta$, $D_s^+ \to \pi^+ \pi^+ \pi^- \pi^0$, $D_s^+ \to \pi^+ \pi^+ \pi^- \pi^0 \pi^0$, $D_s^+ \to K_S^0 K_L^0 \pi^+$, and $D^+ \to K_S^0 K_S^0 \pi^+$. We present results from amplitude analyses revealing rich structures in these decay processes. Furthermore, we report on the $\phi$ decay ratio in $D_s^{\pm}$ decays. Our findings have significant implications for fundamental inputs related to non-perturbative QCD, hadronic matter structure, and studies of $\mathit{CP}$ violation.
}

\FullConference{The European Physical Society Conference on High Energy Physics (EPS-HEP2025)\\
7-11 July 2025\\
Marseille, France\\}

\begin{document}
\maketitle

\section{Introduction}

The study of hadronic decays of charmed mesons (denoted as $D$) serves as a valuable laboratory for investigating both Standard Model dynamics and exotic hadronic phenomena. These decays are primarily characterized by quasi-two-body final states (e.g., $PP$, $VP$, $VV$, $SP$, $AP$), which allow for high experimental precision due to their clean environments and low background noise.

From a theoretical standpoint, these processes offer a crucial testing ground for frameworks such as perturbative QCD (pQCD) through the Topological Diagram Approach~\cite{vp3}. By decomposing amplitudes into color-allowed tree ($T$), color-suppressed tree ($C$), W-exchange ($E$), and W-annihilation ($A$) diagrams, we can extract fundamental parameters of the theory and investigate non-perturbative QCD mechanisms, including final-state interactions, which are notoriously challenging to calculate.

Furthermore, these decays act as a powerful input for hadron spectroscopy. For instance:
\begin{itemize}
  \item \textbf{$D \to PP$ decays:}  
The decay channel $D \to PP$ presents several advantages, including high efficiency, low background, generally high precision, and the absence of interference effects. These characteristics make it an excellent probe for studies of $\mathit{CP}$ violation ($A_{CP}$) and $\eta$–$\eta'$ mixing, among other phenomena.

  \item \textbf{$D \to SP, SV, SS$ decays:}  
  Decays involving scalar mesons are essential for elucidating the long-standing ``scalar meson puzzle,'' helping to determine the internal quark structure of states like $a_0(980)$, $f_0(980)$, $f_0(500)$ and $K_0^*(700)$~\cite{sp1,sp2,sp3}.
  \item \textbf{$D \to VP$ decays:} 
  The well-understood quark content of vector mesons makes channels like $D \to VP$ ideal for isolating and offer cleaner opportunities than other two
body processes for clarifying the non-perturbative mechanism of charmed meson decays~\cite{vp1, vp2, vp3}.
    \item \textbf{$D \to VV$ decays:} 
  The polarization in the $D \to VV$ decay is sensitive to the V-A structure of electroweak interactions, spin correlations, and final state interactions, among other effects~\cite{vv1, vv2}.
  
    \item \textbf{$D \to AP$ decays:} 
  The $D$ meson decays into axial-vector mesons open a window for investigating the properties of states like $a_1(1260)$ and $K_1(1270)$.
\end{itemize}
Thus, precise measurements of these branching fractions~(BFs) and amplitude properties are indispensable for advancing our understanding of QCD, validating theoretical models, and uncovering the nature of light hadrons.

BESIII experiment has collected $e^+e^-$ collision data samples of 20.3 fb$^{-1}$ and 7.33 fb$^{-1}$ at 3.773~GeV and 4.128-4.226~GeV, respectively. These datasets offer the world’s largest samples of charm hadrons near their production threshold, including $D^{+(0)}$ and $D_s^{\pm}$. These datasets offer considerable advantages for amplitude analyses and for measuring BFs with double-tag mathod.

\section{Amplitude analysis of $D^+ \to  K_S^0\pi^+\eta$}

In the study of charmed meson decays, the process \(D^+ \to K_S^0\pi^+\eta\) offers a valuable opportunity to investigate the mechanism of \(D^+ \to a_0(980)^+ P\) decays~\cite{kspieta}. Significant discrepancies have been observed between experimental results and theoretical predictions for the BFs of  \(D^+ \to a_0(980)^+ P\) decays. Among all \(D \to a_0(980)^+ P\) channels, \(D^+ \to K_S^0 a_0(980)^+\) is particularly important, as it is free from weak-annihilation contributions and proceeds predominantly via internal \(W\)-emission. This makes it an ideal process for probing the internal \(W\)-emission mechanism in \(D \to SP\) decays and for providing stringent constraints on topological diagrams within the Standard Model.

The BF of the intermediate process \(D^+ \to K_S^0 a_0(980)^+\) with \(a_0(980)^+ \to \pi^+\eta\) has been measured to be \((1.33 \pm 0.05_{\text{stat}} \pm 0.04_{\text{syst}})\%\). Given the persistent demand from theoretical studies~\cite{th1,th2}, this measurement offers crucial experimental input for validating internal \(W\)-emission contributions and improving the understanding of decay dynamics in charmed meson transitions.

\section{Amplitude analyses of $D^0\to \pi^+\pi^-\eta$ and $D^+\to \pi^+\pi^0\eta$}
We can study $D \to a_0(980)\pi$ decays through the $D^0 \to \pi^+\pi^-\eta$ and $D^+ \to \pi^+\pi^0\eta$ channels~\cite{pipieta}. In $D^0$ decays, both $D^0 \to a_0(980)^+\pi^-$ and $D^0 \to a_0(980)^-\pi^+$ proceed only through $T$ and $E$ diagrams. Theoretical expectations based on isospin symmetry and current models, such as the one proposed by Ref.~\cite{Oset}, predict a small value for the ratio $r_{+/-} = \mathcal{B}(D^0 \to a_0(980)^+\pi^-) / \mathcal{B}(D^0 \to a_0(980)^-\pi^+)$, estimated to be less than 0.05 when $W$-emission contributions are neglected. Previous experimental efforts by the CLEO and LHCb Collaborations to measure this ratio have remained inconclusive due to large uncertainties.

In this analysis, we observe a pronounced enhancement of decays involving the positively charged $a_0(980)^+$. The decay rate for $D^0 \to a_0(980)^+\pi^-$ exceeds that for $D^0 \to a_0(980)^-\pi^+$ by a factor of 7.5. Similarly, $D^+ \to a_0(980)^+\pi^0$ is 2.6 times larger than $D^+ \to a_0(980)^0\pi^+$. These results, which are in clear tension with theoretical predictions by orders of magnitude, suggest the presence of unaccounted mechanisms—possibly within the decay dynamics themselves or in the internal structure of the light scalar mesons, such as the $a_0(980)$.

\section{Amplitude analysis of $D_s^+\to \pi^+\pi^+\pi^-\pi^0$}

The measurements of $D \to SV$ decays are very limited, presenting interesting challenges in $D_s^+\to \pi^+\pi^+\pi^-\pi^0$ decay~\cite{xiaoyu}.
Among $D_{(s)}$ hadronic decays, $D_s^+ \to f_0(980) \rho^+$ is of particular importance as it mainly involves a $W$-external-emission channel, so that final-state interactions are key ingredients in the production of light scalar mesons. These are of particular interest given the lack of consensus on whether these particles are members of the normal scalar meson nonet or four-quark states. Another interesting intermediate decay $D_s^+ \to \phi \pi^+$, which is a key reference channel for $D_s^+$ decays, can be studied through $\phi \to \pi^+ \pi^- \pi^0$. The data in the Particle Data Group~(PDG)~\cite{PDG} on the OZI suppressed $\phi \to \rho \pi$ decays are mostly from $e^+e^-$ collisions on the $\phi$ peak. The possible interference effect between $e^+ e^- \to \gamma^* \to \rho \pi$ and $e^+ e^- \to \phi \to \rho \pi$ may not be well considered, and therefore, BESIII provides another test of the $\phi$ decay, in which no such interference arises.

In the work, observation of $D_s^+ \to f_0(980) \rho(770)^+$ is reported with a statistical significance greater than $10\sigma$. Moreover, the relative BF between $\phi \to \pi^+ \pi^- \pi^0$ and $\phi \to K^+ K^-$ is measured to be
\[\frac{\mathcal{B}(\phi \to \pi^+ \pi^- \pi^0)}{\mathcal{B}(\phi \to K^+ K^-)} = 0.230 \pm 0.014_{\rm stat} \pm 0.010_{\rm syst}.\]
The results significantly deviate from the PDG values obtained in $e^+ e^-$ annihilation and $K^- p$ scattering experiments and may imply the existence of previously unknown interference effects in these processes.

\section{Amplitude analysis of $D_s^+\to \pi^+\pi^+\pi^-\pi^0\pi^0$}
The polarization information of heavy-flavor mesons decaying into two vector particles has attracted the attention of physicists for decades because of its unique advantage in probing new physics and novel phenomena in hadron structures. Among the decays, the comparison between the pure $W$-annihilation decay $D_s^+ \to \omega \rho^+$ and pure $W$-external decay $D_s^+ \to \phi \rho^+$ offers an ideal approach to investigate the mechanism behind the polarization puzzle in $D \to VV$ decays~\cite{zehui}.

In this analysis, the first observation of the pure $W$-annihilation decay $D_s^+ \to \omega \rho^+$, and the anomalous dominance of the $D$ wave over the $S$ and $P$ waves, with a fraction of $(51.85 \pm 7.28_{-7.90}^{+4.83})\%$ are reported. The branching fraction of $D_s^+ \to \phi \rho^+$ is measured to be $(3.98 \pm 0.33_{-0.19}^{+0.21})\%$.

The measured branching fraction provides the first direct experimental determination on a $W$-annihilation process in $D \to VV$ decays. In comparison to the dominance of the $S$ wave and the low significance of the $D$ wave in the pure external $W$-emission decay $D_s^+ \to \phi \rho^+$, the dominance of the $D$ wave in $D_s^+ \to \omega \rho^+$ deviates from the expectation of the naïve factorization model, and is expected to offer important insights for unraveling the ``polarization puzzle''. Furthermore, the ratio $\frac{\mathcal{B}(\phi \to \pi^+ \pi^- \pi^0)}{\mathcal{B}(\phi \to K^+ K^-)}$ is measured to be $0.222 \pm 0.019 \pm 0.016$, which is consistent with the previous measurement~\cite{xiaoyu} in $D_s^+$ mesons but deviates from the PDG value by more than $3\sigma$.

\section{Amplitude analysis of $D_s^+\to K_S^0 K_L^0 \pi^0$}
The $\phi$ meson holds significant importance for studying both the strong and electroweak interactions. Its primary decay channels are the charged channel $\phi \to K^+ K^-$ and the neutral channel $\phi \to K_S^0 K_L^0$, with their combined BFs exceeding 80\%, making them the most crucial decays for investigating the properties of the $\phi$ meson. Thus, precise measurement of these BFs and their ratio $R_{\phi \to K\bar{K}} = \mathcal{B}(\phi \to K_S^0 K_L^0) / \mathcal{B}(\phi \to K^+ K^-)$ would have significant theoretical and experimental implications. However, there are certain inconsistencies between theoretical predictions and experimental measurements of $R_{\phi \to K\bar{K}}$, and it has not been updated for decades. Therefore, there is an urgent need to measure $R_{\phi \to K\bar{K}}$ using a novel approach independent of previous methods, thereby providing experimental basis for relevant theories.

The BESIII collaboration measured the BF of $D_s^+ \to \phi \pi^+$, $\phi \to K_S^0 K_L^0$ through an amplitude analysis of $D_s^+ \to K_S^0 K_L^0 \pi^+$~\cite{huili}. Combining this result with the world average BF of $D_s^+ \to \phi \pi^+$, $\phi \to K^+ K^-$, a precise value of $R_{\phi \to K\bar{K}}$ was determined to be:
\[
R_{\phi \to K\bar{K}} = 0.593 \pm 0.023_{\text{stat}} \pm 0.014_{\text{syst}} \pm 0.016_{\phi\pi},
\]
which deviates from the PDG value by more than \(3\sigma\). This result will have a significant impact on the foundational inputs for studies related to non-perturbative QCD, the structures of hadronic matters, and $\mathit{CP}$ violation. 

The measurement of the $\phi$ meson's decay BF has prompted a series of reflections. For instance, given the currently measured decay proportion of $\phi$ decay, which are significantly less than 100\%, does this suggest the existence of other undiscovered decay channels? Or does it indicate that the current absolute BFs of $\phi$ meson decays need to be remeasured? Furthermore, should existing relevant theories be revised? These questions necessitate further investigation.

\section{Amplitude analysis of $D^+\to K_S^0 K_S^0 \pi^+$}
The $D \to VP$ decays offer cleaner opportunities than other two-body processes for clarifying the non-perturbative mechanism of charmed meson decays. Compared to the $D \to SP$ decay, the $VP$ system presents a more well-defined quark content of the vector meson than that of the scalar state, whose quark content is still a matter of controversy and is further complicated by significant long-distance rescattering effects. The $VP$ system also can be more favorably described by the theoretical framework than the $PP$ and $VV$ systems, due to the multiple-amplitude composition of some $P$ states and the polarization in $VV$ systems.

While most of the theoretical predicted BFs of $D \to VP$ decays are consistent with experimental measurements, that for $D^+ \to K_S^0 K(892)^+$ shows inconsistency. The $D^+ \to K_S^0 K(892)^+$ decay is one of the most important Singly-Cabibbo-suppressed $D \to VP$ decays, which can be mediated via color-favored, W-annihilation, and penguin diagrams.

The BF for $D^+ \to K_S^0 K(892)^+$ is measured to be~\cite{shuai}:
\[
\mathcal{B}(D^+ \to K_S^0 K(892)^+) = (8.72 \pm 0.28_{\text{stat}} \pm 0.15_{\text{syst}}) \times 10^{-3},
\]
which is consistent with the previous BESIII measurement but with improved precision~\cite{2017B}.

\section{Summary}

Charm hadronic decays are key labs to understand non-perturbative QCD. The copious decay products also provide ideal platform to investigate the natures of light mesons, such as $a_0(980)$, $f_0(980)$, $f_0(500)$, $a_1(1260)$, $\phi$ \ldots

Based on $e^+e^-$ annihilation data corresponding to a total integrated luminosity of 7.33fb$^{-1}$ and 7.9fb$^{-1}$ collected at $\sqrt{s} = 4.128-4.226$~GeV and $\sqrt{s} = 3.773$~GeV with the BESIII detector, we report amplitude analyses and the BF measurements for $D_{(s)}$ decay modes. Many structures were observed in these decays and the results are the most precise up to date.
In future, more interesting results are coming using $20.3~\text{fb}^{-1}$ $\psi(3770)$ data.

\section{Acknowledgement}
The author would like to thank all the colleagues in BESIII Collaboration for their great effort to make the mentioned results available, and of course thank the organiser of EPS-HEP 2025 conference to make the extraordinary event
happen. The talk is supported by National Natural Science Foundation of China (Grants No. 123B2077 and No. 12035009).


\begin{thebibliography}{99}
\bibitem{vp3} H.Y. Cheng and C.W. Chiang, \href{https://doi.org/10.1103/PhysRevD.81.074021} {Phys. Rev. D {\bf 81}, 074021 (2010).}
\bibitem{sp1}N. N. Achasov and V. N. Ivanchenko, 
\href{https://linkinghub.elsevier.com/retrieve/pii/0550321389903647} {Nucl. Phys. B, {\bf 315}, 465 (1989). }
\bibitem{sp2}N. N. Achasov and V. N. Ivanchenko, 
\href{https://inspirehep.net/files/cc1210ecc01b82fe2756b3b7582a22e6} {Phys. Lett. B, {\bf 759}, 504 (2016). }
\bibitem{sp3}M. Ablikim {\it et al.}  (BESIII Collaboration)
\href{https://journals.aps.org/prl/abstract/10.1103/PhysRevLett.123.112001} {Phys. Rev. Lett. {\bf 123}, 112001 (2019). }
\bibitem{vp1} H.Y. Cheng, C.W. Chiang, and A.L. Kuo, \href{https://journals.aps.org/prd/abstract/10.1103/PhysRevD.93.114010} {Phys. Rev. D {\bf 93}, 114010 (2016).}
\bibitem{vp2} H.Y. Cheng and C.W. Chiang, \href{https://doi.org/10.1103/PhysRevD.100.093002} {Phys. Rev. D {\bf 100}, 093002 (2019).}

\bibitem{vv1} E. H.E. Aaoud and A.N. Kamal, \href{https://doi.org/10.1103/PhysRevD.59.114013} {Phys. Rev. D {\bf 59}, 114013 (1999).}
\bibitem{vv2} J. Song, Z. Y. Yang, and E. Oset, \href{https://arxiv.org/abs/2501.09422} {arXiv: 2501.09422}

\bibitem{kspieta} M. Ablikim {\it et al.} (BESIII Collaboration), 
\href{https://journals.aps.org/prl/abstract/10.1103/PhysRevLett.132.131903}{Phys. Rev. Lett. {\bf 132}, 131903 (2024).}
\bibitem{th1} H.Y. Cheng, C.W. Chiang, and Z.Q. Zhang, \href{https://doi.org/10.1103/PhysRevD.105.033006} {Phys. Rev. D {\bf 105}, 033006 (2022).}
\bibitem{th2} H.Y. Cheng, C.W. Chiang,
\href{https://doi.org/10.1103/PhysRevD.81.074031} {Phys. Rev. D {\bf 81}, 074031 (2010).}

\bibitem{pipieta} M. Ablikim {\it et al.} (BESIII Collaboration), 
\href{https://journals.aps.org/prd/abstract/10.1103/PhysRevD.110.L111102}{Phys. Rev. D. {\bf 110}, L111102 (2024).}
\bibitem{Oset} M. Ablikim {\it et al.} (BESIII Collaboration), 
\href{https://doi.org/10.1016/j.scib.2023.01.011}{Sci. Bull. {\bf 68}, 243 (2023)}

\bibitem{xiaoyu} M.~Ablikim {\it et al.} (BESIII Collaboration), \href{https://journals.aps.org/prl/abstract/10.1103/PhysRevLett.134.011904}{Phys. Rev. Lett. \textbf{134}, 011904 (2025).}
\bibitem{PDG}S. Navas {\it et al.} (Particle Data Group), 
\href{https://journals.aps.org/prd/abstract/10.1103/PhysRevD.110.030001}{Phys. Rev. D \textbf{ 110}, 030001 (2024).}
\bibitem{zehui} M.~Ablikim {\it et al.} (BESIII Collaboration), \href{https://journals.aps.org/prl/abstract/10.1103/PhysRevLett.134.201902}{Phys. Rev. Lett. \textbf{134}, 201902 (2025).}

\bibitem{huili} M.~Ablikim {\it et al.} (BESIII Collaboration), \href{https://journals.aps.org/prl/abstract/10.1103/6py9-h8qv}{Phys. Rev. Lett. \textbf{135}, 161902 (2025).}

\bibitem{shuai} M.~Ablikim {\it et al.} (BESIII Collaboration), \href{https://journals.aps.org/prd/abstract/10.1103/PhysRevD.110.092006}{Phys. Rev. D. {\bf 110}, 092006 (2024).}

\bibitem{2017B} M.~Ablikim {\it et al.} (BESIII Collaboration), \href{https://doi.org/10.1016/j.physletb.2016.12.020}{Phys. Lett. B. {\bf 765}, 231 (2017).}




\end{thebibliography}
\end{document}